\documentclass[apj,numberedappendix]{emulateapj}
\usepackage{epstopdf}
\usepackage{wrapfig}
\usepackage[utf8]{inputenc}
\usepackage{amsmath}
\usepackage{natbib}
\bibpunct{(}{)}{;}{a}{}{,}

\usepackage{color}
\definecolor{darkgreen}{RGB}{0,142,128}
\definecolor{darkblue}{RGB}{0,100,170}

\usepackage{hyperref}
\hypersetup{colorlinks,citecolor=blue,linkcolor=blue}

\begin{document}
\title{Parametric decay and the origin of the low frequency Alfvénic spectrum of the solar wind}

\author{Victor Réville}
\author{Anna Tenerani}
\author{Marco Velli}
\affil{UCLA Earth, Planetary and Space Sciences, 595 Charles E. Young Drive East, 90095 Los Angeles, CA, vreville@epss.ucla.edu}

\begin{abstract}
The fast solar wind shows a wide spectrum of transverse magnetic and velocity field perturbations. These perturbations are strongly correlated in the sense of Alfvén waves propagating mostly outward, from the Sun to the interplanetary medium. They are likely to be fundamental to the acceleration and the heating of the solar wind. However, the precise origin of the broadband spectrum is to date unknown. Typical periods of chromospheric Alfvén waves are limited to a few minutes, and any longer period perturbations should be strongly reflected at the transition region. In this work, we show that minute long Alfvénic  fluctuations are unstable to the parametric instability. Parametric instability enables an inverse energy cascade by exciting several hours long periods Alfv\'enic fluctuations together with strong density fluctuations (typically between 1 and $20 R_{\odot}$). These results may improve our understanding of the origin of the solar wind turbulent spectrum and will be tested by the Parker Solar Probe.
\end{abstract}


\section{Introduction} 
\label{intro}

Alfvénic perturbations have been observed in the solar wind for almost 50 years \citep{BelcherDavis1971}. Mostly present in fast solar wind streams, they are thought to be involved in the acceleration process of the solar plasma coming from coronal holes. Fast streams are indeed necessarily created through an extended energy deposition \citep[see][]{Leer1982}, which makes the weakly dissipative Alfvén waves a good candidate to transport the energy from the convective motions at the photosphere up to the corona. Observations from the Hinode satellite demonstrated the existence in the chromosphere of Alfvén waves with enough power to drive the solar wind \citep{DePontieu2007}. The wave periods have been characterized between $100$ and $500$ seconds while their amplitudes reached $20$ to $50$ kilometers per second. However, the study of the solar wind magnetic perturbations at large distances shows a broadband power spectrum with frequencies as low as $10^{-4}$-$10^{-5}$ Hz. Moreover, we observe two regimes in the fast solar wind spectrum with a $1/f$ slope at low frequencies that differ from the typical Kolmogorov or Kraichnan slope observed in the higher frequency inertial range \citep[see][and references therein]{BrunoCarbone2013}.

\citet{MatthaeusGoldstein1986,Matthaeus2007} have suggested that the $1/f$ fast wind power spectrum observed at low frequencies could be due to reconnection related magnetic processes occurring in the photosphere and/or the lower corona. Alternatively, it has been shown that the nonlinear interaction between outward and inward fluctuations in the solar wind can give birth to a broadband spectrum \citep{Velli1989,VerdiniVelli2007,Verdini2009,PerezChandran2013}, with the observed slope break \citep{Verdini2012}. The simplest mechanism to generate these inward waves is reflection. In the incompressible limit, MHD equations predict that any large scale gradient in the solar wind speed and in the Alfvén speed will reflect forward Alfvén waves coming from the Sun, reflection becoming stronger for decreasing frequency.  A frequency dependent transmission coefficient can thus be studied to characterize their propagation \citep[see][for a review]{Velli1991}. Significant reflection occurs for low frequency waves with periods over a few hours, which are dominant in the corona, hence providing the necessary ingredients for an incompressible solar wind turbulence. However, at the transition region, the very sharp gradient of the Alfvén speed should prevent hour long fluctuations from freely propagating into the low corona. Some early studies have indeed shown that the transmitted power from the photosphere to the corona would present resonant peaks that would transmit very little power upward \citep{Hollweg1972,Hollweg1978a,Leroy1981}. \citet{Velli1993} later remarked that these peaks could be in principle be removed using a more regular Alfvén speed profile (more precisely $\mathcal{C}^1$, \textit{i.e.} with a continuous derivative), making the transition region essentially a high pass filter (see Appendix \ref{app:B} for an extended discussion).

The question therefore remains: how do low frequency waves reach the corona and how do they come to dominate the solar wind spectrum given that higher frequencies should be more easily transmitted through the transition region? One possibility is to assume that the longer periods are created by an inverse cascade process in the low corona. This process should be able to reconstruct the solar wind power spectrum from the transmitted waves typically observed in the chromosphere. The parametric decay instability (PDI) is a credible candidate, creating from a mother forward Alfvén wave two daughter waves: a forward acoustic wave and an inward Alfvén wave at a longer wavelength. This compressible process has been studied extensively from a theoretical point of view \citep[see][]{GaleevOraevskii1963,Goldstein1978,Derby1978,JayantiHollweg1993,MalaraVelli1996,Chandran2018} and within numerical simulations \citep{Malara2000,DelZanna2001,Tanaka2007,TeneraniVelli2013,Shi2017}. While PDI is undoubtedly an important process in low beta plasmas, it is still a matter of debate whether the dynamics of the solar wind, namely the expansion and the acceleration \citep[or even kinetic effects, see e.g,][for 3D hybrid simulations]{Fu2018}, prevent this process to be a significant ingredient of solar wind turbulence \citep[see][]{TeneraniVelli2013}.

The recent study of \citet{Shoda2018b} has tackled the onset of the parametric decay instability in a realistic wind profile. They find large density perturbations, which they assume to be due to the instability, to develop for frequencies higher than $10^{-3}$ Hz. In this paper, we adopt a similar approach, including the transition region, hence accounting for the reflection of low frequency waves. We characterize precisely the onset of the instability for $f \geq 2 \times 10^{-3}$ Hz. We further demonstrate that the PDI is first responsible for the creation of an inward wave and an inverse cascade, yielding a broadband perturbation spectrum in a solar wind solution. The parametric decay instability relies fundamentally on acoustic perturbations, and we solve the fully compressible MHD equations in a single flux tube, while injecting Alfvén waves from the photosphere and through the transition region. In Section \ref{sec:num}, we describe the numerics of our model and the steady state atmosphere capturing the transition region. In Section \ref{sec:mono}, we analyze the propagation of monochromatic, circularly polarized waves from the chromosphere to the corona. We show that low frequency waves ($f \leq 10^{-3}$ Hz) are systematically filtered (through an almost total reflection). Then, we precisely characterize the parametric decay occurring for frequencies $f>10^{-3}$ Hz inside the domain and we show that the growth rate is in general in agreement with theory. We further consider in Section \ref{sec:spectra} non-monochromatic sources of constant $||B_{\perp}||$. We show that the parametric decay process is not altered by a broadband spectrum injection, if not enhanced in some cases. Finally, we describe the inverse and forward cascade occurring after the non linear evolution of the Alfvén and acoustic waves in Section \ref{sec:cascade}. We discuss the implications, limits and perspectives of our findings in Section \ref{sec:disc}.

\section{Numerical setup}
\label{sec:num}

\begin{figure*}
\centering
\includegraphics[width=7in]{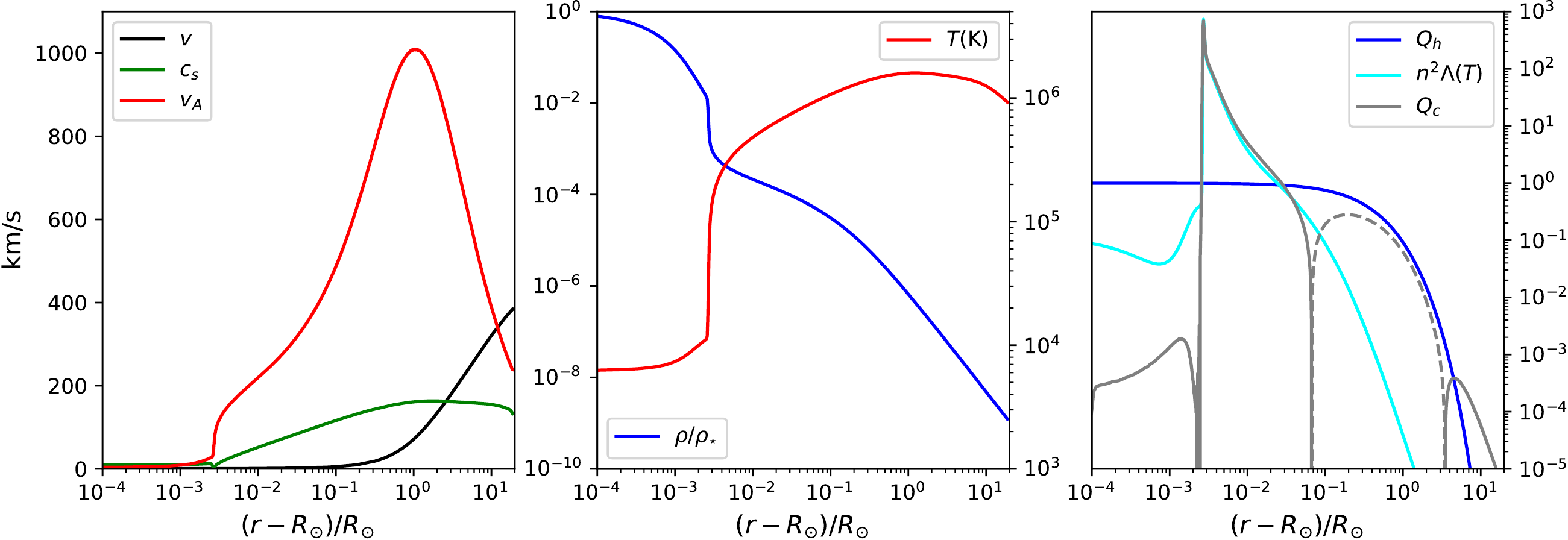}
\caption{Steady-state atmosphere. On the left panel we show the outflow speed $v$, the sound speed $c_s$ and the Alfvén speed $v_A$. The middle panel shows the temperature and density profiles and the right panel the balance of heating, radiation and thermal conduction of equation \ref{eq:q} in units of $F_h$ (negative when dashed).}
\label{fig:SteadyWind}
\end{figure*}

\subsection{MHD model}
We solve the time dependent ideal magnetohydrodynamics (MHD) equations using the PLUTO code \citep{Mignone2007}: 
\begin{equation}
\label{MHD_1}
\frac{\partial}{\partial t} \rho + \nabla \cdot \rho \mathbf{v} = 0,
\end{equation}
\begin{equation}
\label{MHD_2}
\frac{\partial}{\partial t} \rho \mathbf{v} + \nabla \cdot (\rho \mathbf{vv}-\mathbf{BB}+\mathbf{I}p) = - \rho \nabla \Phi,
\end{equation}
\begin{equation}
\label{MHD_3}
\frac{\partial}{\partial t} (E + \rho \Phi)  + \nabla \cdot ((E+p+\rho \Phi)\mathbf{v}-\mathbf{B}(\mathbf{v} \cdot \mathbf{B})) = Q,
\end{equation}
\begin{equation}
\label{MHD_4}
\frac{\partial}{\partial t} \mathbf{B} + \nabla \cdot (\mathbf{vB}-\mathbf{Bv})=0,
\end{equation}
where $E \equiv  \rho e + \rho v^2/2 + B^2/2$ is the total energy, $\mathbf{B}$ is the magnetic field, $\rho$ is the mass density, $\mathbf{v}$ is the velocity field, $p = p_{\mathrm{th}}+B^2/2$ is the total (thermal plus magnetic) pressure and $\mathbf{I}$ is the identity matrix. The vectors fields have three components along the radial, poloidal and azimuthal directions $(r,\theta,\varphi)$, but depend only on $r$, the distance to the Sun, describing the evolution of a single, radial flux tube under a gravity potential

\begin{equation}
\Phi = - \frac{GM_{\odot}}{r}.
\end{equation}

The source term $Q$ added to the energy equation is made of three components:
\begin{equation}
\label{eq:q}
Q = Q_h - Q_r - Q_c,
\end{equation}
where the usual heating, cooling and thermal conduction sources are implemented as follows:

\begin{equation}
Q_h = F_h/H \left(\frac{R_{\odot}}{r} \right)^2 \exp{ \left(-\frac{r-R_{\odot}}{H}\right)},
\end{equation}
with $H=1 R_{\odot}$, the heating scale-height, and $F_h = 1.5 \times 10^5$ erg.cm$^{-2}$s$^{-1}$ the energy flux from the photosphere (see section \ref{sec:scm}). We then used an optically thin radiation cooling prescription,

\begin{equation}
Q_r = n^2 \Lambda (T),
\end{equation}
with $n$ the electron density and $T$ the electron temperature. $\Lambda(T)$ is defined as in \citet{Athay1986}. The thermal conduction flux combines a collisional and a collisionless prescription :

\begin{equation}
Q_c = \nabla \cdot (\alpha \mathbf{q}_s + (1-\alpha) \mathbf{q}_p),
\end{equation}
$\mathbf{q}_s$ being the usual Spitzer-Härm collisional thermal conduction with $\kappa_0=9 \times 10^{-7}$ cgs, and $\mathbf{q}_p = 3/2 p_{\mathrm{th}} \mathbf{v}$ the free-stream heat flux \citep{Hollweg1986}. The coefficient $\alpha = 1/(1+(r-R_{\odot})^4/(r_{\mathrm{coll}}-R_{\odot})^4)$ creates a smooth transition between the two regimes at a characteristic height of $r_{\mathrm{coll}} = 5 R_{\odot}$. Finally, an ideal closure equation relates the internal energy and the thermal pressure, 

\begin{equation}
\rho e  = \frac{p_{\mathrm{th}}}{\gamma-1},
\end{equation}
with $\gamma = 5/3$, the ratio of specific heat for what we consider a fully ionized hydrogen gas. 

The equations are solved using an improved Harten, Lax, van Leer Riemann solver \citep[HLLD, see][]{MiyoshiKusano2005}, combined with a parabolic reconstruction method and minmod slope limiter. We maintain $\nabla \cdot \mathbf{B} = 0$ using a hyperbolic divergence cleaning method \citep{Dedner2002}. 

\subsection{Steady Atmospheric Structure}
\label{sec:scm}

In this specific study we limit our domain to $[1 R_{\odot}, 20 R_{\odot}]$, discretized with 16384 grid points. A first set of 128 cells are used to describe the domain up to $1.001 R_{\odot}$ at very high resolution. Then, a stretched grid of 2048 cells is used to progressively decrease the resolution up to $1.5 R_{\odot}$. The last grid is uniform and can, with a resolution of $\sim 10^{-3}R_{\odot}$, accurately describe the propagation of the waves up to the upper boundary for all frequencies. In a first step, we evolve the system keeping zero transverse velocity and magnetic field. The steady coronal structure is shown in figure \ref{fig:SteadyWind}. The lower boundary is alike a photosphere/chromosphere, \textit{i.e.} a stratified atmosphere at a fixed temperature $T= 6000K$, close to hydrostatic equilibrium. The phenomenological heating term $Q_h$ heats the atmosphere, up to a maximum temperature around $1.7$ MK. The atmosphere thus sustains a transition region located between $10^{-3} R_{\odot}$ and $10^{-2} R_{\odot}$ above the solar surface. MK temperatures then drive a wind that becomes supersonic at $r_c = 3.6 R_{\odot}$ and superalfvénic at $r_A =  13 R_{\odot}$. 

The heating rate $F_h$ has been chosen to obtain a mass loss consistent with observations, $\dot{M} = 3 \times 10^{-14} M_{\odot}$/yr. Its value is somewhat lower that what can be found in other works, because we neglect the superradial expansion of the flux tube (see Appendix \ref{app:A}). The magnetic field hence decays as $1/r^2$  and the surface field $B_{\odot} = 1.5$ G. The base density is $n =  10^{12}$ cm$^{-3}$. The wind speed is around $380$ km/s at the edge of the domain. This value varies only little with the value of $F_h$ chosen, but has a strong positive dependence on the scale height.  The value of $H=1R_\odot$ is typical of such single flux tube setup \citep[see][]{Pinto2009,Grappin2010}, and can represent various heating sources in the low corona: acoustic heating and shock dissipation or resistive heating through reconnection processes such as nanoflares \citep{Parker1988}. This wind solution is typical of a slow wind with relatively low speed and mass loss on the upper range of the observed values. In the following, we use this solution as an equilibrium background to propagate Alfvén waves, which provide an additional source of heating and momentum to produce a faster wind \citep[see, e.g.,][for a specific study on the terminal wind speed and comparison with in-situ data at 1 AU]{Shoda2018a}.

\section{Alfvén wave propagation, reflection and parametric decay}
\label{sec:mono}
\subsection{Monochromatic wave injection}

\begin{figure}
\includegraphics[width=3.2in]{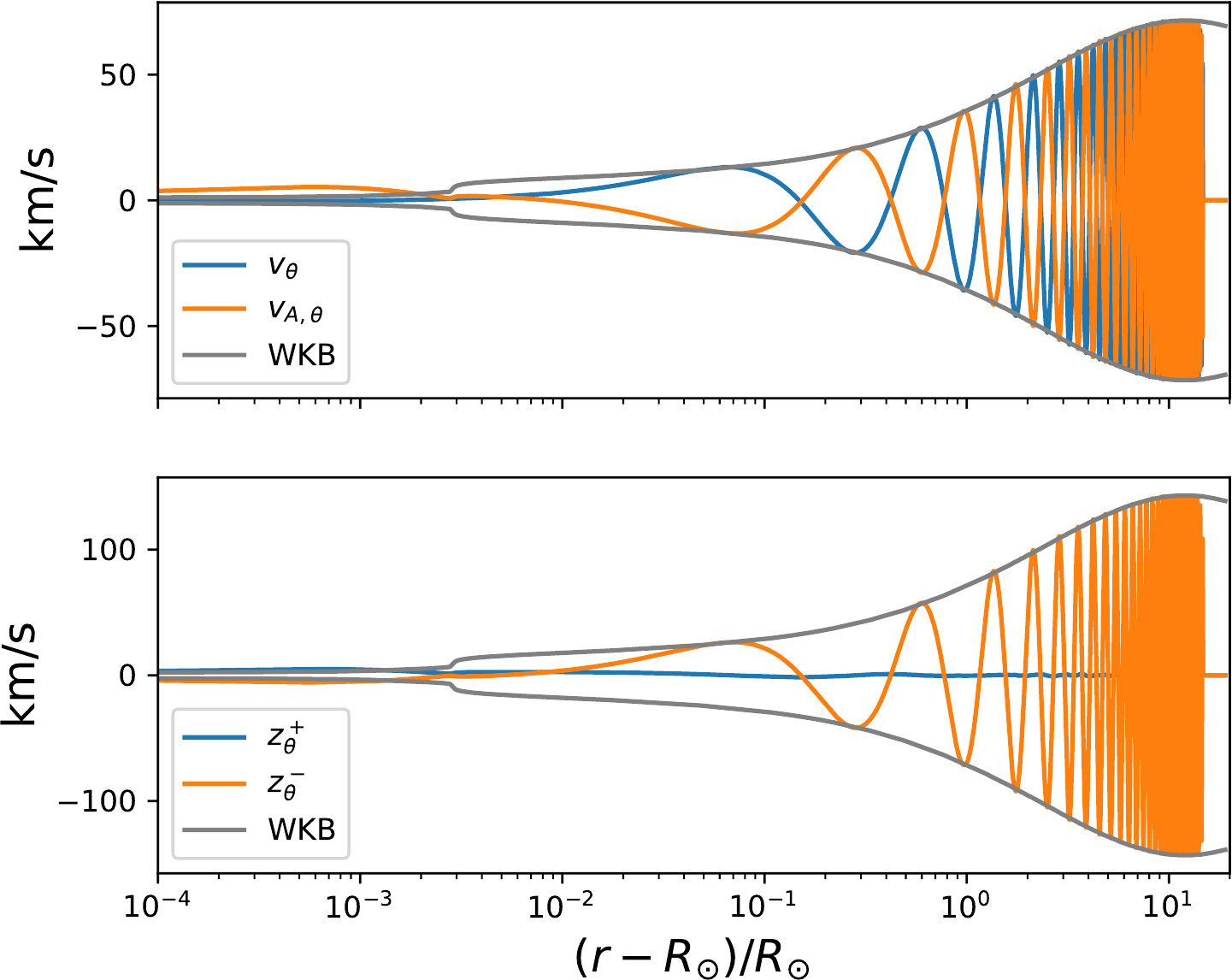}
\caption{Propagation of a circularly polarized Alfvén wave of frequency $2 \times 10^{-3}$ Hz and base amplitude $2$ km/s at $t=0.7 t_A$, before the onset of the parametric decay. In the top panel we show the azimuthal velocity and magnetic field (through the Alfvén velocity) that are in phase opposition. We display the theoretical WKB envelope in gray. In the bottom panel we see the outwardly propagating waves $z^-$ with no sign of reflected wave $z^+$ outside the transition region.}
\label{fig:prop}
\end{figure}

We now inject circularly polarized Alfvén waves at the low boundary of the domain. These waves are known to be hard to dissipate and can be as such seen as the remnants of the dynamic forcing at the photosphere, which are able to go through the transition region and the lower corona, as shall be seen further. We define the Elsässer variables:

\begin{equation}
\mathbf{z}^{\pm} = \mathbf{v}_{\perp} \pm \frac{\mathbf{b}_{\perp}}{\sqrt{\mu_0 \rho}},
\end{equation}

corresponding to fluctuations that propagate parallel (lower sign, in our case propagating outwards) or antiparallel (upper sign, in our case propagating inwards) to the background magnetic field, and we impose the outwards propagating Alfv\'en wave from the lower boundary as:
\begin{equation}
\mathbf{z}^- = 2 |\delta v| (\cos(\omega_0 t) \mathbf{e}_{\theta} + \sin(\omega_0 t) \mathbf{e}_{\varphi}),
\end{equation}
where $\omega_0 = 2\pi f_0$ is the input pulsation.  The amplitude and frequency are typical of what is observed both in the photosphere/chromosphere and in the lower corona. We consider $|\delta v| \in [1, 2]$ km/s and a range of periods $P=1/f_0 \in [50, 100, 200, 500, 1000, 10000]$ seconds. The pump wave is forced from the inner boundary condition throughout the simulation.

In Figure \ref{fig:prop}, we show the propagation of the Alfvén wave inside the computational domain at the Alfvén speed. We define

\begin{equation}
t_A = \int_{R_\odot}^{20 R_{\odot}} \frac{dr}{v+v_A} = 18072 \;\mathrm{s},
\end{equation}
the time for Alfvén waves to cross the computational domain. In Figure \ref{fig:prop}, $t=0.7 t_A$, and the input wave as propagated beyond $r=14R_{\odot}$. The amplitude is around $25$ km/s at $r=1.5 R_{\odot}$. In the bottom panel of Figure \ref{fig:prop}, we can see a significant $z^+$ component below the transition region, almost equal in amplitude to $z^-$. This component is created by reflections on the very sharp gradient of the Alfvén speed at the transition region ($r-R_{\odot} \approx 5 \times 10^{-3}$, see Figure \ref{fig:SteadyWind}). We impose a zero gradient boundary condition on reflected incoming waves at the lower boundary $\partial z^+/\partial r = 0$, while other quantities are maintained at their equilibrium value. The transition region acts as a fully self-consistent inner boundary for the transmitted waves. Outside the transition region, we see the amplitude of the Alfvén waves growing with distance accordingly to the Wentzel–Kramers–Brillouin (WKB) theory \citep[][for which we used the base of the corona as a reference value]{Parker1965,Belcher1971}. A weak $z^+$ inward component is necessarily present as a result of the continuous reflections in the corona, but appears to have a negligible amplitude \citep[see][for an analytical derivation with a two-scale approach and Section \ref{sec:tr} and \ref{sec:spectra}]{TeneraniVelli2017}. For the outer boundary we use characteristic boundary conditions to avoid non physical reflections as the outward waves escape the domain \citep[see the appendix of][]{Landi2005}.

\subsection{Reflections in the transition region}
\label{sec:tr}
The amount of reflected wave on the steep gradient of $v_A$ at the transition region is a function of frequency. Defining a proper transmission coefficient in a stratified atmosphere is non trivial as a reference state with constant Alfvén speed is in general required \citep[see][]{Velli1991}. We can however use the conservation of wave action \citep{BrethertonGarret1968,Jacques1977} to quantify the wave energy that is able to reach the corona \citep[see][for a generalization of this result]{HeinemannOlbert1980,Chandran2015}. We define the wave action flux
\begin{equation}
S^{\pm} = \frac{(v\mp v_A)^2}{v_A} \rho r^2 \frac{|z^{\pm}|^2}{8},
\end{equation} 
for the forward and inward Alfvén waves. In the absence of non linear interactions, the total wave action is conserved, which means that  $S^--S^+=\;$ constant in steady state \citep[see e.g.][]{VerdiniVelli2007}. As shown in Figure \ref{fig:prop}, past the transition region the forward wave amplitude follows  the WKB profile, meaning that $S^+$ is negligible and $S^-$ is constant in the corona.

\begin{figure}
\includegraphics[width=3.2in]{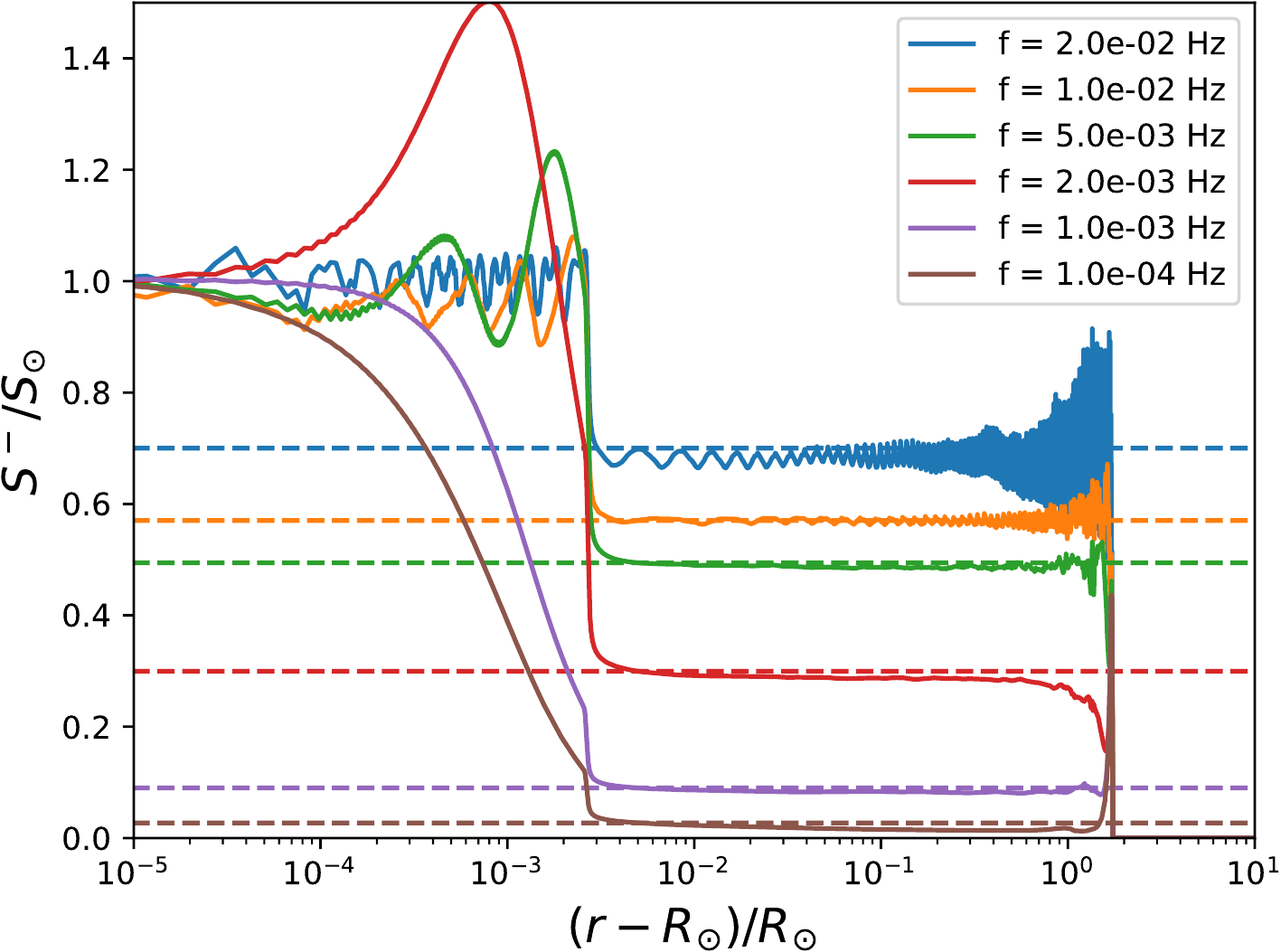}
\caption{Outward wave action flux at $t=0.1 t_A$ for different frequencies and $\delta v = 1$km/s. Outside the transition region the wave action flux is roughly constant which allow to define an transmission coefficient $\mathcal{T}$, shown with the dashed lines. Signs of parametric decay start to be visible at the highest frequencies.}
\label{fig:ActT}
\end{figure}

\begin{deluxetable}{l|c|l|l|l|l}
\tablecaption{Case parameters and results}
\tablecolumns{4}
\tabletypesize{\scriptsize}
\tablehead{
	\colhead{$f_0$ (Hz)} &
    \colhead{$\delta v$ (km/s)} &
    \colhead{$\mathcal{T}$}&
	\colhead{$t_{\mathrm{on}}\; (t_A)$} &
	\colhead{$\gamma_p$ (s$^{-1}$)} &
    \colhead{$\gamma_{\mathrm{th}}$ (s$^{-1}$)}
	 }
	\startdata
	$2 \times 10^{-2}$ & 1.0 & 0.75  & 0.73 & 2.5e-3 & 4.2e-3   \\
	$1 \times 10^{-2}$ & 1.0 & 0.57  & 0.81 & 1.0e-3 & 3.2e-3    \\
 	$5 \times 10^{-3}$ & 1.0 & 0.48  & 1.32 & 1.0e-3 & 1.5e-3  \\
	$2 \times 10^{-3}$ & 1.0 & 0.30  & -    & -      & -       \\
    $1 \times 10^{-3}$ & 1.0 & 0.09  & -    & -      & -       \\
    $1 \times 10^{-4}$ & 1.0 & 0.027 & -    & -      & -       \\
    $2 \times 10^{-2}$ & 2.0 & 0.67  & 0.14 & 9.0e-3 &  1.0e-2 \\
	$1 \times 10^{-2}$ & 2.0 & 0.57  & 0.27 & 4.2e-3 &  5.0e-3 \\
 	$5 \times 10^{-3}$ & 2.0 & 0.49  & 0.57 & 2.0e-3 &  3.0e-3 \\
	$2 \times 10^{-3}$ & 2.0 & 0.30  & 1.37 & 8.0e-4 &  8.1e-4 \\
    $1 \times 10^{-3}$ & 2.0 & 0.09  & -    & -      & -       \\
    $1 \times 10^{-4}$ & 2.0 & 0.027 & -    & -      & - 
	\enddata
\tablecomments{Parameters and characteristics of monochromatic cases: the transmission coefficient $\mathcal{T}$, the time of the instability onset $t_{\mathrm{on}}$, the simulation and theoretical growth rates $\gamma_p$, $\gamma_{\mathrm{th}}$.}
\label{table1}
\end{deluxetable}

In Figure \ref{fig:ActT}, we plot the evolution of the outward wave action flux in the domain for different frequencies. Beyond the transition region, $S^-$ reaches a plateau, which defines the transmission coefficient of the forward Alfvén waves $\mathcal{T}$. The values are reported in Table \ref{table1}. Waves are more and more reflected as their period increases, and $\mathcal{T}$ is below $10 \%$ for $f_0 \leq 10^{-3}$ Hz, around $3 \%$ for $f_0 = 10^{-4}$ Hz.

In Table \ref{table1} we see that the transmission coefficient varies strongly with frequency, but very little with amplitude, at least for the parameter range chosen. Many models have tried to give an analytical estimate for $\mathcal{T}$, we review and compare some of them with our results in Appendix \ref{app:B}. One reliable feature in all these models is that the chromosphere and the transition region behave as a high pass filter for Alfvén waves, which is verified with our simulations. Low frequency waves are strongly reflected and hardly reach the corona. Hence, their existence in the solar wind may be an indicator of inverse cascade processes. 

\subsection{Parametric decay: resonance and growth rate}

We observe the development of parametric decay in a subset of our simulations. In our model, the plasma beta parameter, which we define as the squared ratio of the sound speed over the Alfvén speed $\beta = c_s^2/v_A^2$, is of the order $0.1$ in the corona and the relative perturbations $\delta b/B \approx 0.1$. In this regime, Alfvén waves are known to be unstable to parametric decay. However, the onset of the instability on time scales shorter or comparable to the Alfvén crossing time is a function of the wave frequency and the wave amplitude in the corona, itself modified by the frequency dependent transmission through the transition region. Wind expansion and acceleration also act to suppress the instability and, for the present parameter space we find the onset threshold of the instability to be around $f_0 = 2 \times 10^{-3}$ Hz. At this frequency, we observe the onset of the instability for $\delta v = 2$ km/s, while the case $\delta v = 1$ km/s remains stable. For all higher frequencies, the instability grows for all base amplitude. For $f_0 \leq 10^{-3}$ Hz we only observe weak reflection driven inward waves without non linear interactions for at least $3 t_A$.

\begin{figure}
\center
\includegraphics[width=3.2in]{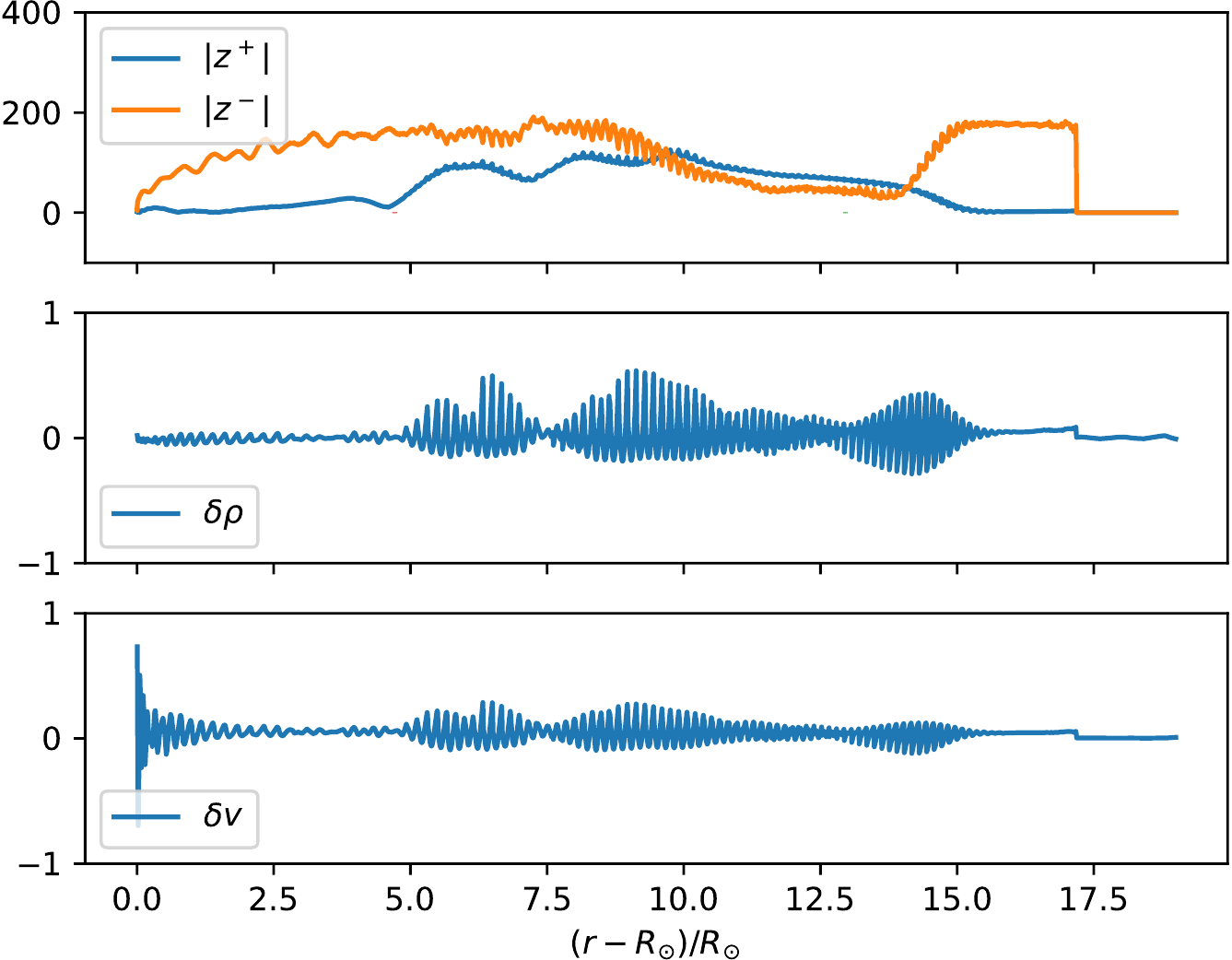}
\caption{Profiles of the parametric instability onset phase for $f=5 \times 10^{-3}$ Hz and $\delta v = 2$ km/s. Wave packet of density and radial velocity perturbations form in multiple locations in the domain and create an inward $z^+$ propagating Alfvén wave (shown in km/s). The forward wave has not yet reached the outer boundary of the domain ($t = 0.88 t_A$).}
\label{fig:ParProfile}
\end{figure}

In Figure \ref{fig:ParProfile}, we illustrate a typical case of the parametric decay instability growth phase. The top panel shows the forward ($z^-$) and inward ($z^+$) Alfvén waves. The pump forward Alfvén waves is propagating into the computational domain ($t=0.85 \; t_A$) and we observe within the domain density and radial velocity perturbation in the middle panel and bottom panel respectively with:
\begin{equation}
\delta \rho = \frac{\rho-\rho_0}{\rho_0},\; \delta v = \frac{v-v_0}{v_0},
\end{equation}
the $0$ subscript corresponding to the steady state solution described in \ref{sec:scm}. It is worth noting that the instability is triggered in multiple regions at the same time. \citet{Tanaka2007} have shown similar structure for the perturbations that were located in one region only. We observe wave packets of various periods and an inverse cascade characterized by the longer wavelength modulations. The growth region of density and velocity perturbations can clearly be associated with the creation of an inward Alfvén wave and the corresponding decay of the pump forward waves (see Figure \ref{fig:ParProfile}, top panel). 

To perform a  more quantitative characterization of the instability we rely on the resonance conditions between the outward propagating mother Alfvén waves, the daughter inward Alfvén wave and outward sound wave. We must have :

\begin{equation}
f_{A,\mathrm{out}} = f_S + f_{A,\mathrm{in}},
\end{equation}
and

\begin{equation}
\mathbf{k}_{A,\mathrm{out}} = \mathbf{k}_S + \mathbf{k}_{A,\mathrm{in}}.
\end{equation}

\begin{figure}
\includegraphics[width=3.4in]{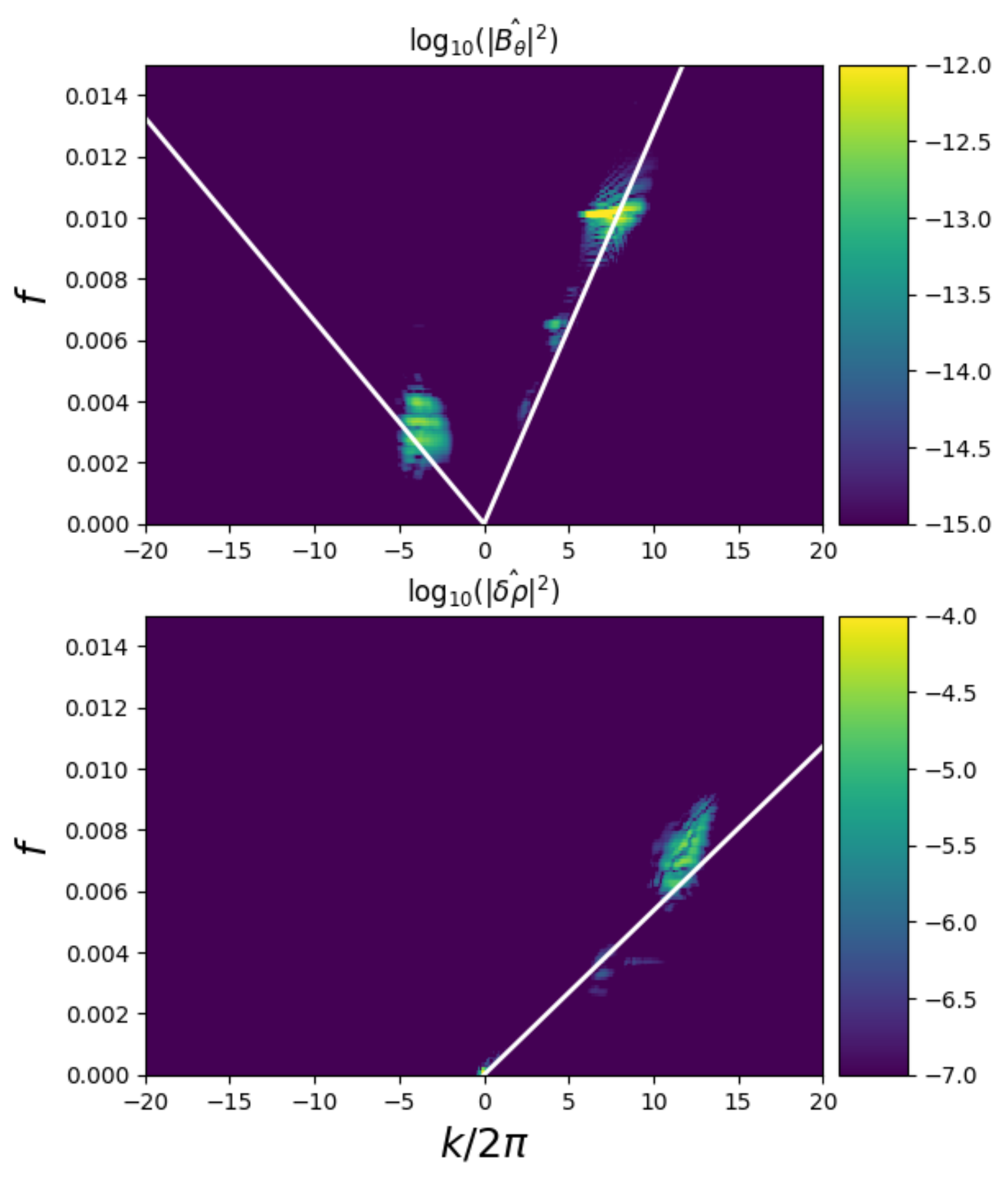}
\caption{Fourier transform of the transverse magnetic field and the density variation as a function of $f$ and $\bar{k} \equiv k /2 \pi$. The top panel show the pump Alfvén mode at $\bar{k}=7 R_{\odot}^{-1}$ , $f = 10^{-2}$ Hz, and the reflected Alfvén wave at $\bar{k}=-5 R_{\odot}^{-1}$ , $f = 3 \times 10^{-3}$ Hz. The averaged dispersion relations  $f=(\langle v \rangle \pm \langle v_A\rangle) \bar{k}$ are shown in white. In the bottom panel the forward sound wave appear clearly along the curve $f=(\langle v \rangle + \langle c_s \rangle) \bar{k}$ at $\bar{k} = 12 R_{\odot}^{-1}$. The resonance condition for the parametric instability is verified in space and time.}
\label{fig:Omk}
\end{figure}

In Figure \ref{fig:Omk}, we illustrate these resonance conditions for $f_0=10^{-2}$ Hz, $\delta v = 2$ km/s, using a 2D discrete Fourier transform for the growing phase of the instability \citep[see][]{Tanaka2007}. In the top panel we show the $(f,\bar{k} \equiv k/2\pi)$ spectrum of one of the perpendicular component of the magnetic field. The spectrum is obtained considering the growth region (here between $1.5 R_{\odot}$ and $7 R_{\odot}$) and the growth time span (between $0$ and $1 t_A$) for this specific case. The power is located at two main points, the pump forward and the daughter inward Alfvén waves at $\bar{k}=7 R_{\odot}^{-1}, f= 10^{-2}$ Hz and $\bar{k}=-5 R_{\odot}^{-1}, f= 3 \times 10^{-3}$ Hz respectively. The white lines represent the approximate dispersion relation $f=(\langle v \rangle \pm \langle v_A \rangle) \bar{k}$, where brackets denote the spatial average taken over the growth domain of the perturbations. As can be seen, the pump and resonant modes are in excellent agreement with the dispersion relation. 

Similarly, in the bottom panel of Figure \ref{fig:Omk}, we plot the $(f,\bar{k})$ spectrum of the density perturbations $\delta \rho$\footnote{The analysis of the radial velocity perturbations yields the same results.}. The curve $f = (\langle v \rangle + \langle c_s \rangle) \bar{k}$, where 
\begin{equation}
c_s = \sqrt{\frac{\partial p}{\partial \rho}},
\end{equation}
is the effective sound speed, is shown again in white. The power is concentrated on a forward sound wave at $\bar{k}=12 R_{\odot}^{-1}, f= 7 \times 10^{-3}$ Hz. Here, and in all the cases reported in Table \ref{table1}, the resonance conditions are symptomatic of the parametric decay occurring in the simulation. This process is the dominant and first non linear process breaking the WKB approximation, giving birth to an inward Alfvén wave.

\begin{figure}
\includegraphics[width=3.2in]{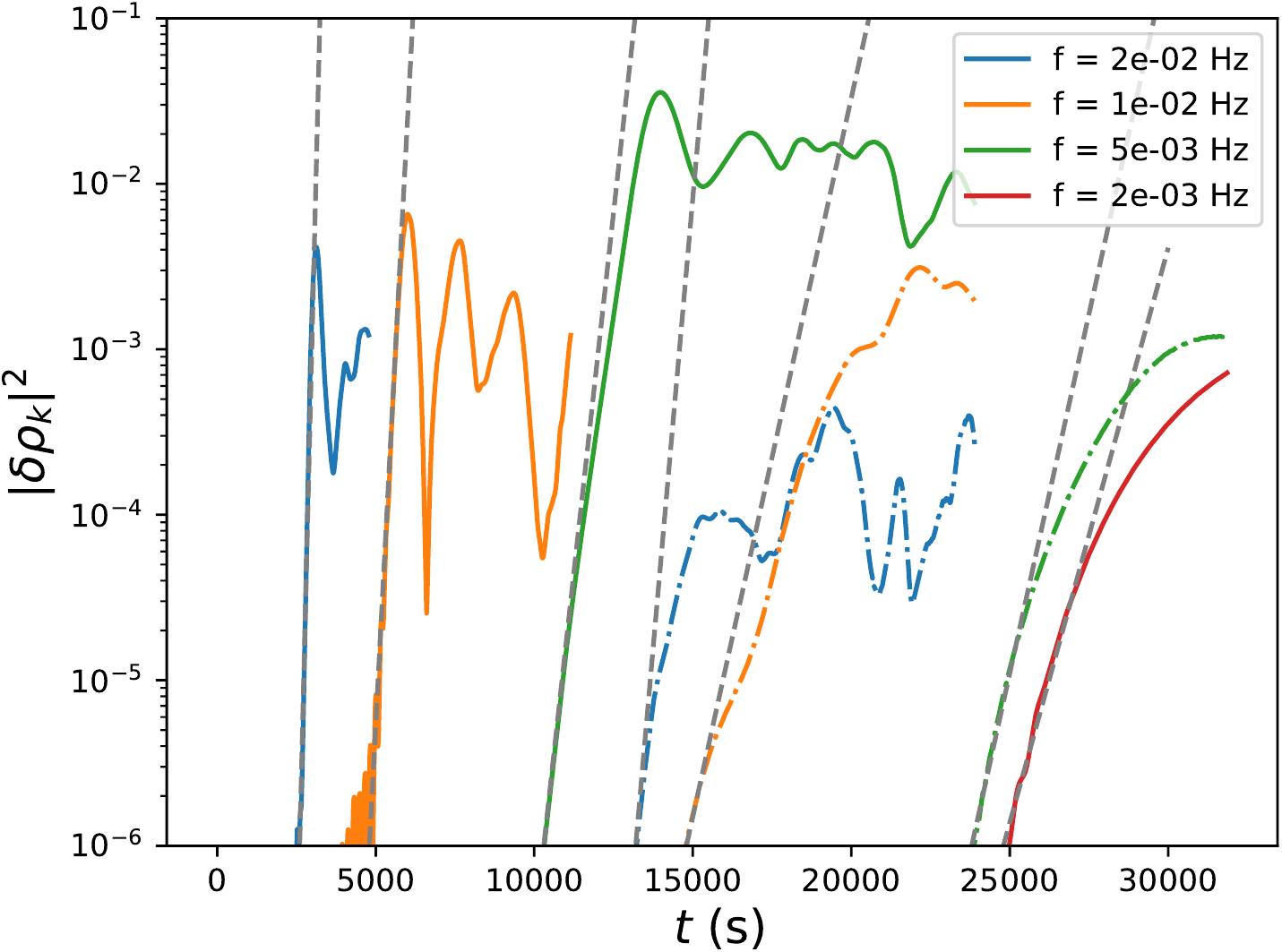}
\caption{Growth of the acoustic mode associated with the parametric instability. Plain lines are for $\delta v = 2$ km/s and dot-dashed lines for $\delta v = 1$ km/s. The dashed gray lines fit equation (\ref{eq:growthfit}) with $A = 10^{-6}$ and the values reported in Table \ref{table1} for $t_{\mathrm{on}}$ and $\gamma_p$.}
\label{fig:growth}
\end{figure}

We go further in Figure \ref{fig:growth}, where we show the growth of the forward acoustic mode for different simulations.  For all unstable cases, the acoustic mode follows an exponential phase led by a saturation phase. We fit the exponential phase with the law :

\begin{equation}
|\delta \rho_k|^2 (t) = A \exp(2\gamma_p (t-t_{\mathrm{on}})),
\label{eq:growthfit}
\end{equation}
and report the result in Table \ref{table1}. The growth rate $\gamma_p$ is an increasing function of the frequency and scales almost linearly with $f_0$, especially for $\delta v =2$ km/s. We note also that the growth rate is naturally lower for lower base amplitudes. The onset time $t_{\mathrm{on}}$ follows the opposite trend, \textit{i.e} is a decreasing function of the growth rate. 

We can compare the value obtained in Table \ref{table1}, with analytical estimates. We first consider the classical homogeneous dispersion relation \citep{Goldstein1978,Derby1978} :

\begin{equation}
(\omega-k)(\omega^2-b^2 k^2) [(\omega+k)^2-4] = a^2 k^2 (\omega^3+k\omega^2 - 3\omega+k),
\end{equation}
where $a=\delta b/B$ and $b=\sqrt{\beta}=c_s/v_A$ and $\omega$ and $k$ are normalized by the pump frequency and wavenumber. From \citet{JayantiHollweg1993} we have in the case of small $a$ an estimate for the maximum growth rate
\begin{equation}
\gamma_{\mathrm{max}} = \omega_0 \frac{a(1-b)^{1/2}}{2b^{1/2}(1+b)}.
\label{eq:gest}
\end{equation}
where we immediately notice the linear relation between $\gamma_{\mathrm{est}}$ and $f_0$ and the wave amplitude. We note also that when $b > 1$, \textit{i.e} $\beta > 1$, the system is stable. In our steady-state atmosphere the sound speed crosses the Alfvén speed beyond $20 R_{\odot}$ and thus all the domain is potentially unstable to parametric decay.

Following \citet{TeneraniVelli2013,Shoda2018b}, we correct this estimate taking into account the effect of the expansion and of the acceleration of the solar wind, we define 
\begin{equation}
\gamma_{\mathrm{th}} = \gamma_{\mathrm{max}} - \gamma_{\mathrm{acc}} - \gamma_{\mathrm{exp}},
\label{eq:gth}
\end{equation}
where 

\begin{equation}
\gamma_{\mathrm{acc}} = \frac{\partial}{\partial r} (v_{r,0} + c_s),
\end{equation}
and

\begin{equation}
\gamma_{\mathrm{exp}} = (v_{r,0}+c_s) \frac{\partial}{\partial r} \ln A = (v_{r,0}+c_s) \frac{2}{r},
\end{equation}
where $A$ is the area of the flux tube, here purely radial. 

\begin{figure}
\includegraphics[width=3.2in]{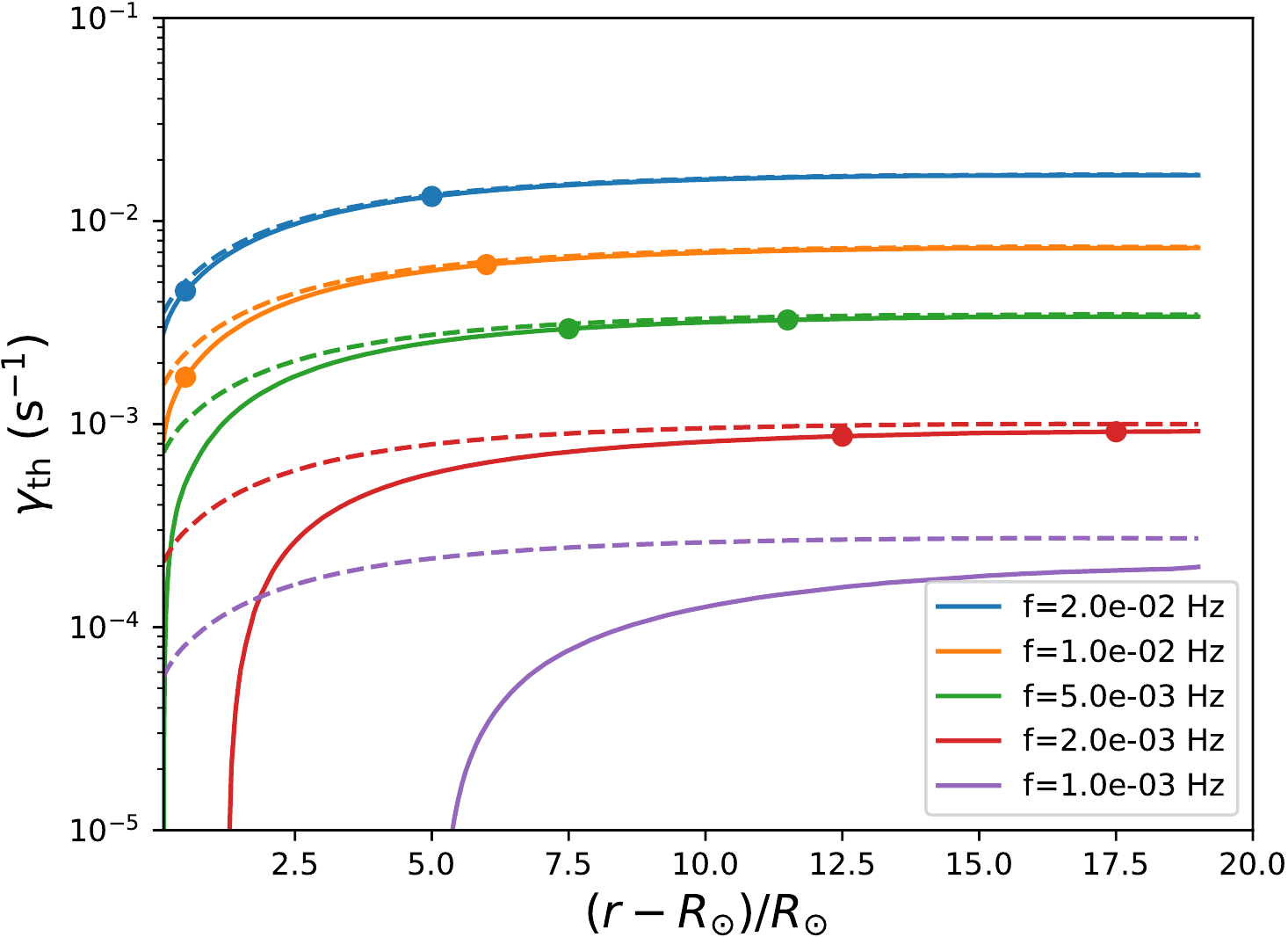}
\caption{Theoretical estimate of the growth rate for the daughter sound wave $\gamma_{\mathrm{th}}$ for various frequencies and $|\delta v|=2$ km/s. The dashed line corresponds to the estimate without taking into account the expansion and acceleration of the wind. For the wind profile considered in this study, the theory gives an unstable $f=10^{-3}$ Hz case and a stable $f=10^{-4}$ Hz. Our study shows that both are stable.}
\label{fig:gam_eff}
\end{figure}

Figure \ref{fig:gam_eff} shows the value of the estimate $\gamma_{\mathrm{th}}$ as a function of the radial distance for various periods. The profile of $b=c_s/v_A$ is computed using the steady solution of section \ref{sec:scm} while $\delta b$ in $a$ is computed using the WKB approximation from the value outside the transition region (see Figure \ref{fig:prop}). Hence, the transmission coefficient $\mathcal{T}$ is fully taken into account. The value $\gamma_{\mathrm{th}}$ reported in Table \ref{table1} is averaged over a sub-domain where the instability develops in each simulation, and shown between the two dots in Figure \ref{fig:gam_eff}. This sub-domain is also used to compute the Fourier transform in space and extract the simulation growth rate $\gamma_p$. We see an overall good agreement between the theoretical values and those obtained with the simulations, peculiarly for $\delta v = 2$km/s. It is worth noting here that the instability develops further from the Sun at lower frequencies, increasing the averaged value of the theoretical growth rate estimate\footnote{The growth rate is an increasing function of the distance from the Sun as can be seen in Figure \ref{fig:gam_eff}.}. This effect seems to compensate the smaller transmission coefficient for lower frequency waves, and allow to conserve the linear relationship between $\gamma_p$ and $f_0$ (see equation (\ref{eq:gest})) for $\delta v = 2$km/s.

Our estimation of the effect of the expansion and acceleration of the solar wind is independent of the frequency and amplitude and lower $f_0$ and $\delta v$ are more sensitive to these effects. This could explain the lower growth rate we obtain for $\delta v = 1$ km/s compared to the maximum theoretical estimate. Also, the theory gives a positive value for $\gamma_{\mathrm{th}}$ when $f_0=10^{-3}$ Hz, while we found this case to be stable in our simulations for all amplitudes. More theoretical developments, beyond the linear analysis resulting in equation (\ref{eq:gth}), might be necessary to accurately describe the transition to stability.

\section{Weakly non monochromatic and broadband spectra}
\label{sec:spectra}

We further consider small and large deviations to the monochromatic wave injection. We use a similar formulation to the one introduced in \citet{MalaraVelli1996} and \citet{TeneraniVelli2013}. The phase of the injected wave is given by:

\begin{equation}
\phi (f_0,t) = 2\pi f_0 t + \varepsilon \sum_{n=2}^{N} n^{-\alpha} \cos (2\pi n f_0 t + \phi_n),
\label{eq:rand}
\end{equation}
where $\phi_n \in [0, 2\pi]$ is a random phase and $\varepsilon$ controls the deviation from a monochromatic wave. $N=512$ typically sets the width of the spectrum, here about 3 decades above the fundamental frequency $f_0$. From equation (\ref{eq:rand}), it can be seen that the average of $\phi(f_0,t)$ remains at the fundamental frequency. The spectral slope of the wave created is a decreasing function of $\alpha$, although the precise value depends on the properties of the random numbers $\phi_n$ and of $\varepsilon$ \citep[see][]{MalaraVelli1996}. Only playing with the phase allows to maintain the circular polarization of the pump Alfvén wave.

\begin{figure}
\includegraphics[width=3.2in]{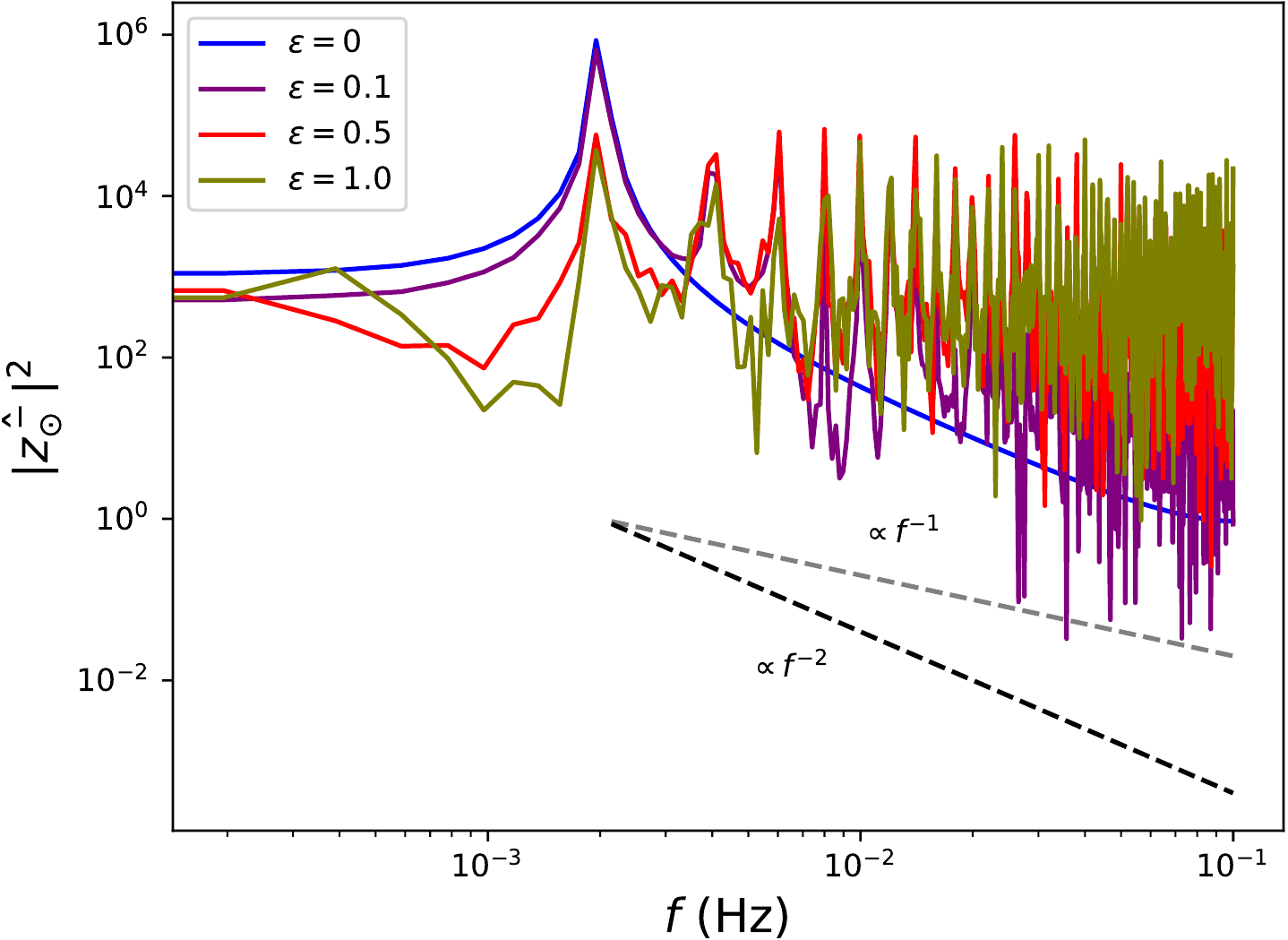}
\caption{Power spectrum (in arbitrary units) of the pump forward wave, whose phase in given by equation (\ref{eq:rand}). Here we chose $f_0 = 2 \times 10^{-3}$ Hz. The spectral slope lies between $f^{0}$ and $f^{-2}$ depending on the value of $\varepsilon$.}
\label{fig:randspectra}
\end{figure}

Figure \ref{fig:randspectra} shows typical spectra of the pump wave at the base of the domain with $\varepsilon \in [0,0.1,0.5,1.0]$ (with $\varepsilon = 0$ corresponding to the monochromatic case), $f_0 = 2 \times 10^{-3}$ Hz, $\alpha = 1$. As we depart from monochromatic waves the power spectrum becomes flatter and flatter and the power in the fundamental frequency $f_0$ diminishes. For $\varepsilon > 0.1$ the power spectrum of input perturbation is well developed with a slope close to $1/f$.

\begin{figure}
\includegraphics[width=3.2in]{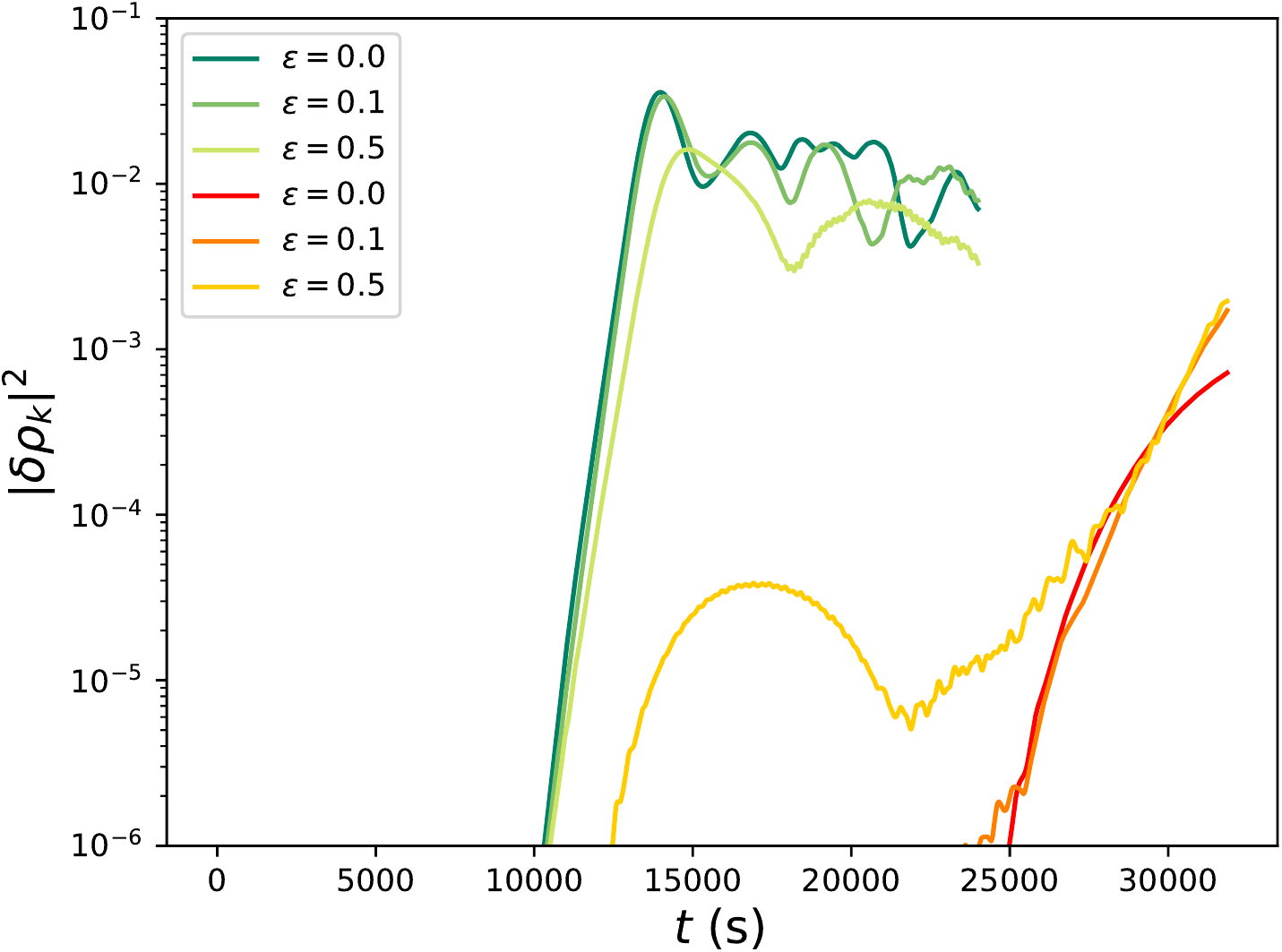}
\caption{Growth of the unstable acoustic modes for $\varepsilon = 0, 0.1, 0.5$ and $f_0 = 5 \times 10^{-3}$ Hz (green variations), $f_0=2 \times 10^{-3}$ Hz (red variations). Here $\delta v =2 $ km/s. The growing non monochromatic parameter has no significant influence for the well established unstable case (green), while it somewhat enhances the growth of the instability for threshold frequencies (red), thanks to higher frequency forcing.}
\label{fig:randgrowth}
\end{figure}

In this section, we consider $\alpha = 1$ and two frequencies, one for which the instability is clearly well developed in the monochromatic case $f_0 = 5 \times 10^{-3}$ Hz, and the second being near the threshold of the onset $f_0 = 2 \times 10^{-3}$ Hz  (we have seen that the system is decaying for $\delta v = 2$ km/s but not in the case of $\delta v = 1$ km/s when the wave is monochromatic).  In Figure \ref{fig:randgrowth}, we plot the growth rate of the unstable acoustic mode for $\varepsilon = 0, 0.1, 0.5$ for those two cases (one can verify that cases with $\varepsilon=0$ correspond with Figure \ref{fig:growth}).

The greenish curves represent the first case. When $\varepsilon$ increases, the instability still onsets and develops at roughly the same time. However the growth rate is slightly smaller for larger $\varepsilon$. This result has been first obtained in \citet{MalaraVelli1996}, where it is shown that the difference between the growth rate of the monochromatic and the one with the phase formulation (\ref{eq:rand}) is a positive value, $\mathcal{O}(\varepsilon^2)$. Hence, for high frequency average forcing, the parametric instability is only marginally affected by a non monochromatic input. 

When $\varepsilon > 0$, higher frequency forward Alfvén waves coexist in the simulation and can trigger the instability. For a weakly non monochromatic input ($\varepsilon = 0.1$, orange curve), the first growing acoustic mode is the same as the monochromatic case. The instability shows a first slow growth phase that is triggered slightly before the monochromatic exponential growth. As shown in \citet{Malara2000}, this first phase can be explained by dissipation, here due to the numerical scheme. The acoustic mode then experiences a exponential growth with a rate similar to the monochromatic cases but with higher levels of saturation. For larger deviations ($\varepsilon = 0.5$, yellow curve), it is a higher frequency wave ($f \sim 2f_0$) that first launches the instability, before the onset time of the monochromatic case. The growth rate is at first higher than in the monochromatic case and eventually reaches a similar level. Hence, we see that in all our cases, a broadband input spectrum does not prevent the instability to develop and for some cases near the stability threshold, it can accelerate the onset and enhance the growth rate.

Nevertheless, we find that the monochromatic study is a good proxy to characterize the presence of the parametric decay. Although we observe small density perturbations for the case $\varepsilon=0.1$, $f_0=2\times 10^{-3}$ Hz and $\delta v = 1$km/s, they are not enough to trigger an exponential phase and do not yield inward Alfvén daughter waves. For lower frequencies, the instability is suppressed and the threshold for the instability onset remains unchanged when $\varepsilon > 0$.

\section{Non linear evolution: forward and inverse cascade}
\label{sec:cascade}

\begin{figure*}
\center
\includegraphics[width=7in]{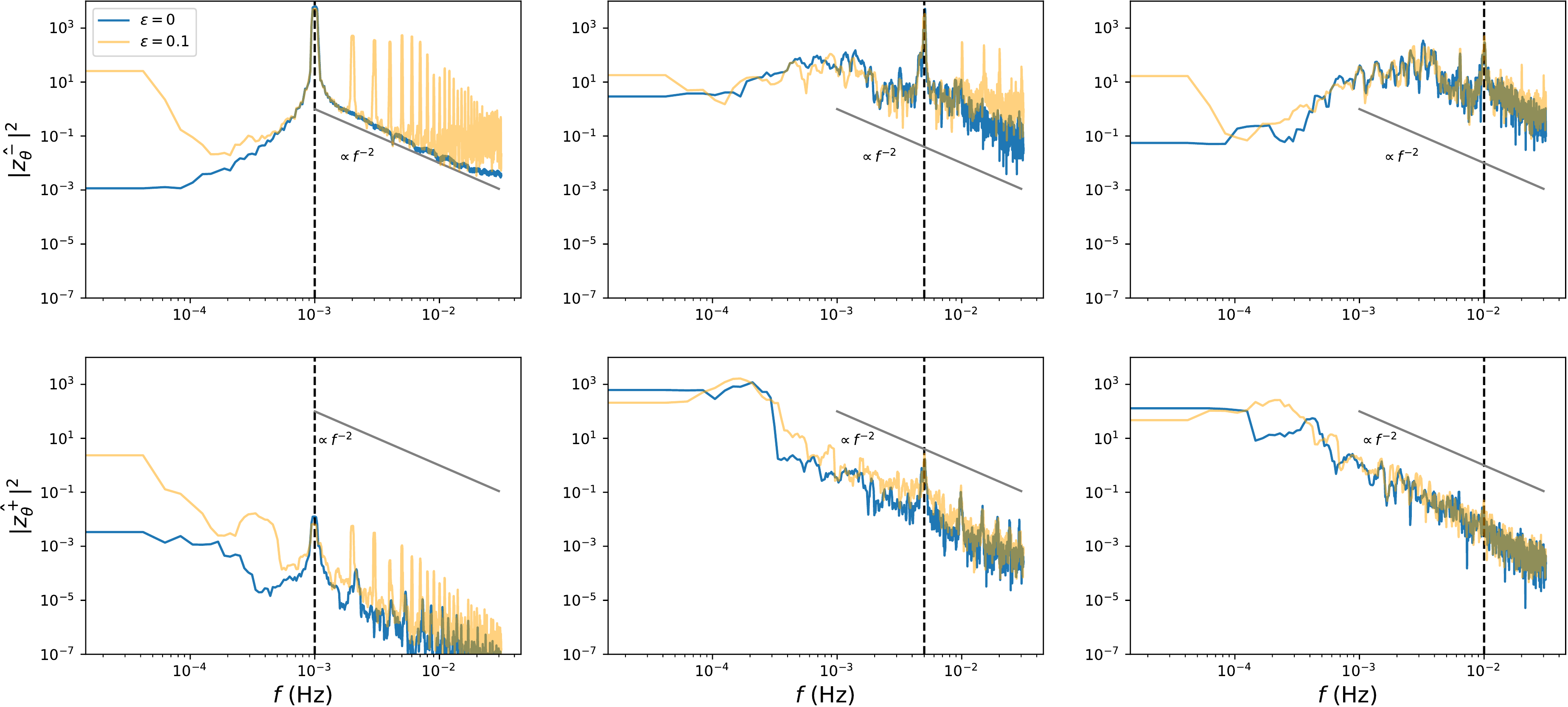}
\caption{Power spectrum of $z^-_{\theta}$ and $z^+_{\theta}$ at $10 R_{\odot}$ (arbitrary units). The left, middle and right panel correspond to frequencies $f_0 = 10^{-3}$ Hz, $5 \times 10^{-3}$ Hz and $10^{-2}$ Hz respectively (shown by the black dashed lines). The faint yellow curves use a weakly non monochromatic pump wave ($\varepsilon = 0.1$). In both cases, the inverse cascade is clear for $f > 10^{-3}$ Hz, where the parametric instability develops. The left panel gives a reference of a stable pump wave with a weak reflection driven inward component. Gray areas represent the superposition of blue and yellow curves.}
\label{fig:spectra}
\end{figure*}

As noted in section \ref{sec:mono}, the parametric decay is able to provide an inverse cascade from the pump waves frequency to lower ones. In this section, we take a closer look at the effect of the onset of the parametric instability on the power spectrum of the solar wind. We plot in Figure \ref{fig:spectra}, the power spectrum at $10 R_{\odot}$, of the forward $z^-$ and inward $z^+$ component of Alfvénic perturbations for three different frequencies. The spectra have been computed over a time range of $3 t_A$, letting time for non linear interactions to evolve. The monochromatic case is shown in plain blue while a similar case with $\varepsilon = 0.1$ is shown in faint yellow.

We first discuss the monochromatic cases ($\varepsilon = 0$, blue curves). In the left panel, the instability has not developed. The forward wave ($z^-$, top panel) spectrum is essentially a single peak at the input frequency. The bottom panel confirms that a small amplitude inward wave is created by continuous reflections as an order one process \citep[with a small parameter $\epsilon = dv_A/dr/\omega_0 \sim 5\times 10^{-3}$, see][]{Velli1989,TeneraniVelli2017}. The inward wave ($z^+$) maximum peak at $f_0 = 10^{-3}$ Hz is more than 4 orders of magnitude smaller than for the two other input frequencies shown in the middle and right panel, where the parametric decay has been triggered. We can thus say that, at least for this range of frequencies where reflections are weak in the corona \citep{Velli1991}, the parametric instability is a much more efficient process to create a significant inward component.

On the middle and right panel, with input frequencies unstable to PDI, we observe the forward and inverse cascade process occurring for $z^+$. It is striking to see that the maxima of the inward Alfvén waves are well below the input frequency, namely at $1.9 \times 10^{-4}$ Hz and $1.6 \times 10^{-4}$ Hz for $f_0 = 5 \times 10^{-3}$ Hz, $f_0 = 10^{-2}$ Hz respectively. The non-linear interaction between $z^+$ and $z^-$ is visible through the resonant peaks beyond the input frequency (particularly in the middle panel). The spectral slopes of the inward component are very close to a $f^{-2}$ law, consistent with what has been predicted in the theoretical work of \citet{Chandran2018}. 

Consequently, low frequencies are recovered in the forward wave spectra (upper panels) with comparable power to the one found at the input frequency, which remain the first maximum. For $f_0 = 5 \times 10^{-3}$ Hz, the secondary maximum of the $z^-$ spectrum is around $2 \times 10^{-4}$ Hz, while for $f_0 = 10^{-2}$ Hz, the secondary maximum is around $3 \times 10^{-3}$ Hz. The inverse cascade is however covering a broad range of frequencies and with an input period of $100$ seconds, we excite periods up to $30$ minutes, while with an input period of $200$ seconds the inverse cascade significantly excites periods up to $3$ hours. We notice a consequent change of slope that is occurring at the input frequencies or, in other words, at the limit between the forward and the inverse cascade. We fitted the forward cascade spectral slopes for the two unstable cases and found spectral indices of $-2.2$ and $-2.3$ for the middle and right panel respectively. They are close to $f^{-2}$ as shown in Figure \ref{fig:spectra}.

When $\varepsilon = 0.1$, \textit{i.e} for a weakly non monochromatic input, the spectral slopes of the forward cascade of the forward wave ($z^-$) are enhanced with indices of $-1.5$ and $-1.8$ for the middle and right panels respectively. This is very likely coming from the input signal as shown in Figure \ref{fig:randspectra}. However, the dynamics created by the parametric instability, namely the creation of the inward wave and the inverse cascade remain globally unchanged. 

\section{Discussion}
\label{sec:disc}

In this study, we have shown that the parametric decay instability is triggered in the low corona ($1-20 R_{\odot}$) when high frequency ($\geq 2 \times 10^{-3}$ Hz) Alfvén waves are launched from the photosphere. These waves, which do not have to be monochromatic, but only to have a significant power at high frequency, are then able to generate low frequency ($\sim$ hour long) perturbations in the solar atmosphere through an inverse cascade process. This mechanism may be the reason for the presence of hour long periods in the solar wind perturbation spectrum. We show indeed that these low frequencies are strongly reflected at the transition region.

The precise values of the transmission coefficient and of the frequency stability threshold can certainly vary with the adopted background solar wind model. Specifically, the addition of coronal superradial expansion would amplify the wave amplitude and possibly displace (to lower values) the stability frequency limit, here at $2 \times 10^{-3}$ Hz). Also, the exponential heating prescription, which shapes the Alfvén speed profile at the transition region, is an external source and is consequently not fully self-consistent. 
 
The model of \citet{Shoda2018b} includes a coronal superradial expansion ($f_{\mathrm{tot}} = 10$) and turbulence induced heating. They find the parametric instability to onset at a similar -although lower- threshold  ($10^{-3}$ Hz). This slight difference may be explained by our modeling of the transition region that significantly reflects low frequency modes near $10^{-3}$ Hz and makes their amplitude in the corona too low to trigger the instability and/or the expansion accounted for in their model. Nevertheless, both studies show that several hundreds of seconds periods, \textit{i.e.} typical frequencies of observed chromospheric Alfvén waves, will very likely trigger the decay despite the expansion and the acceleration of the solar wind.

Another, perhaps more critical, limit of our model is the flux tube approximation that implies a purely parallel Alfvén wave propagation. The study of \citet{SimilonZargham1992}, using an axisymmetric equilibrium model, has shown that a very large value of the expansion factor ($f_{\mathrm{tot}} = 400$), associated to the expansion from the intergranular network, could smooth the Alfvén speed profile to obtain a more or less flat transmission coefficient profile. More realistic configurations, where open regions emerge from complex and multipolar magnetic structure, could also strongly enhance the dissipation and affect the propagation of waves, through phase mixing for instance \citep{HeyvaertsPriest1983,Pucci2014}. Multidimensional simulations are needed to understand the contributions of all these effects \citep[see e.g.,][]{MatsumotoSuzuki2012} and are considered as a follow up of this paper.

Although the spectra shown in Figure \ref{fig:spectra} are the result of the non linear evolution of the system, it is important to stress that PDI creates an inward Alfvén wave much more efficiently than reflection, which is very small, at high frequency. The PDI seems also necessary for the onset of the inverse cascade. These results suggest that the generation of solar wind turbulence is made in-situ in the low corona and that this process must be compressible. Sign of strong density perturbations have been recently observed in the very low corona by \citet{Hahn2018}, where they are identified as possible indicators of the parametric instability. \citet{Chandran2018} has shown that the PDI can be responsible for the $1/f$ slope of the fast wind perturbation spectrum, in a model that damps acoustic perturbation immediately. So far, observations have shown that down to $0.3$ AU, the density perturbations are very weak in the fast wind (as well as the inward component), although signs of local parametric decay have been inferred from statistical study of the solar wind at 1 AU \citep{Bowen2018}. In our simulations, when a pseudo steady state is reached after the onset of parametric decay, we observe density perturbations $\delta \rho$ going from $0.5$ to $0.1$ when moving outward from $5$ to $20 R_{\odot}$, showing that they are already damped within the MHD framework, likely through shock dissipation. Moreover, as we go further out and that the plasma becomes less and less collisional, kinetic processes such as Landau damping will help make vanish density perturbations.

Parametric instability also destroys the Alfvénic correlation of the outward mode. If parametric instability is actually occurring in the low corona, then some other process acts to create or recreate this correlation in the expanding wind. This will be directly tested in-situ by Parker Solar Probe, whose orbit will go down to $9.8 R_{\odot}$ while continuously measuring plasma parameters and electromagnetic fields.

\section{Acknowledgements}
The authors are grateful to A. Mignone and the PLUTO development team. This research was supported by the NASA Parker Solar Probe Observatory Scientist grant NNX15AF34G. This work used the Extreme Science and Engineering Discovery Environment \citep[XSEDE,][]{Xsede2014}, which is supported by National Science Foundation grant number ACI-1548562, through the SDSC based resource Comet with allocation number AST180022. This study has made use of the NASA Astrophysics Data System.

\appendix
\section{Minimum Poynting flux for an expanding wind}
\label{app:A}
Considering a single flux tube allows to make an energy budget between arbitrary surfaces crossed by this flux tube \citep[see e.g.][]{Suzuki2006}. From the total energy conservation equation we have in steady state: 

\begin{equation}
\nabla \cdot \left[ \rho \mathbf{v} ( 1/2 v^2 + \frac{c_s^2}{\gamma-1} -\frac{GM_{\odot}}{r})  + \mathbf{F}_h + \mathbf{F}_c \right] + Q_r = 0,
\end{equation}
where $\nabla \cdot \mathbf{F}_h =  -Q_h$ and $\nabla \cdot \mathbf{F}_c =  Q_c$  are the Poynting and the thermal conduction flux. Considering $A_0$ a surface at the photosphere where we neglect everything but the Poynting flux and the gravity potential, and a surface $A_1$ at  where we neglect everything but the kinetic bulk energy of the wind, we are left with:
\begin{equation}
F_h A_0 - \rho_0 v_0 A_0 \frac{GM_{\odot}}{R_{\odot} }=  A_1 \rho_1 v_1  \frac{1}{2} v_1^2 + \int_{A_0}^{A_1} Q_r dV
\end{equation}

For the wind to escape we must have $v_1 \sim v_{\mathrm{esc}} = \sqrt{2GM_{\odot}/R_{\odot}}$. Using mass conservation and given that $Q_r \geq 0$ we have

\begin{equation}
F_{h} \geq \frac{A_1}{A_0} \rho_1 v_{\mathrm{esc}}^3 = f_{\mathrm{tot}} \left (\frac{r_1}{r_0}\right)^2 \rho_1 v_{\mathrm{esc}}^3.
\end{equation}

We see that the minimum Poynting flux is linearly related to the total expansion factor $f_{\mathrm{tot}}$. For typical values solar wind parameters at earth orbit, and $f_{\mathrm{tot}} \approx 10$ \citep[see e.g.][]{Pinto2016,Reville2017}, the minimum Poynting flux is around $10^{6}$ erg.cm$^{-2}$.s$^{-1}$. In this study, where $f_{\mathrm{tot}} = 1$, the Poynting flux should be above $10^{5}$ erg.cm$^{-2}$.s$^{-1}$. Computing \textit{a posteriori} the radiative losses in our steady-state solution, we find that they account for about $1/3$ of the input flux. The value we chose for our simulations $F_h = 1.5 \times 10^5$ erg.cm$^{-2}$.s$^{-1}$ is thus fully coherent with this analysis.

\section{Analytical computations of the transmission coefficient through the transition region}
\label{app:B}

In this appendix, we discuss some models of the transmission coefficient through the transition region. We rely mostly on the model of \citet{Leroy1981}. In this model, the lower layers of the solar atmosphere are described using a bi-exponential profile of the density corresponding to two stratified isothermal regions at $T_1$ and $T_2$ and a transition region at a height $h$. Two characteristic scale heights,  $H_1$ and $H_2$, are defined for each isothermal layer.

\begin{figure*}
\center
\includegraphics[width=5.5in]{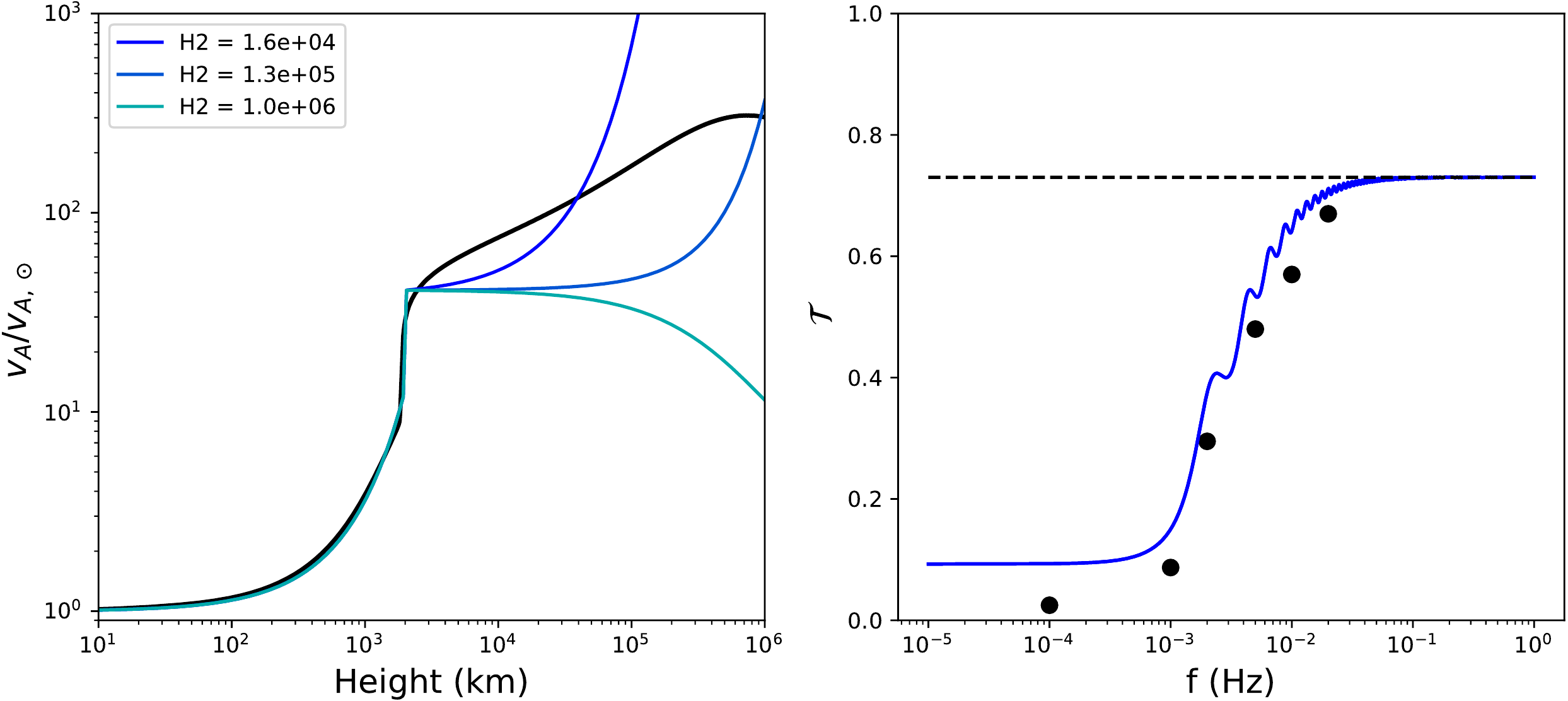}
\caption{Alfvén speed profiles given by the bi-exponential approximation of \citet{Leroy1981} for different values of $H_2$ (left panel). The black line is the steady profile of our study. The corresponding transmission coefficient $\mathcal{T}$ is shown on the right panel. $\mathcal{T}$ is computed at the end of the transition region hence does not depend on $H_2$. The asymptotic value for high frequencies is shown in dashed black. Values from our simulations are scattered in black (see Table \ref{table1})}
\label{fig:Trans}
\end{figure*}

The height of the transition region and the first scale height $H_1$ can be easily constrained to model our chromosphere, the density being very close to a hydrostatic equilibrium profile at $T_1=6000 $ K before the transition region. In Figure \ref{fig:Trans}, we show the profile of the Alfvén speed of our simulations, and several analytical profiles for increasing values of $H_2$. It is clear that the second exponential cannot be a reasonable approximation for the solar corona, even for very small distance beyond the transition region. This is key to what follows.

\citet{Hollweg1972,Hollweg1978a} has first discussed the problem of the transmission of Alfvén waves into the corona through a sharp transition region. He found peaks of transmitted power for "resonant" frequencies. Following the work of \citep{Leroy1981}, which displayed similar resonances, a debate rose to determine whether these were due to the exponential profiles or to the discontinuity at the transition region. In Figure \ref{fig:Trans}, we show the values of $\mathcal{T}$ obtained with our simulations. There are little reasons to believe that these values do not follow a somewhat continuous, smooth profile. The blue profile of the left panel shows a solution of the model of \citet{Leroy1981}, obtained nonetheless with some subtleties. By computing this coefficient at the top of the transition region, we do not take into account the second scale height but only the density jump at the transition region ($\propto T_1/T_2$, where $T_2 = 60000$ K). The transmission profile is considerably smoothed compared to the results displayed in \citet{Leroy1981} where $\mathcal{T}$ was computed at a height of $3 R_{\odot}$. This profile fit fairly well the values we obtain in the simulations. 

This indicates that a large part of the resonances are due to the exponential profiles used to model the corona in both \citet{Hollweg1978a} and \citet{Leroy1981}. Moreover, as shown in Figure  \ref{fig:prop} and \ref{fig:ActT}, the inward Alfvén wave is negligible before the onset of the parametric decay instability. This is equivalent to say that the corona is at least for some time a transparent medium for Alfvén waves. The blue profile in Figure \ref{fig:Trans} keeps however some oscillatory features that are likely related to the discontinuity in scale heights (or derivative) of the density profile of the model, as shown by \citet{Velli1993}. 

Resonant peaks are thus probably pure artifacts of the analytical models. The general behavior that remains is that low frequencies ($< 10^{-3}$ Hz) are strongly reflected, even more so than what is predicted by the model of \citet{Leroy1981}, which loses some validity for long periods \citep[see the numerical solutions of][]{SimilonZargham1992,VerdiniVelli2007}. Hence, the argument of \citet{Hollweg1978a} that hour long Alfvén waves are transmitted through resonances seems unlikely and the most reliable feature of the transition region is to be a high pass filter when a significant gradient of Alfvén speed is present. However, in the presence of very large expansion around the transition region, which acts to smooth the Alfvén speed, the transmission coefficient could loose this property and yield a flat profile around 30\% \citep{SimilonZargham1992}.

\end{document}